\documentclass[pdflatex,sn-mathphys]{sn-jnl} 
\jyear{2021}
\theoremstyle{thmstyleone}

\theoremstyle{thmstyletwo}%

\theoremstyle{thmstylethree}%

\begin{document}

\title[Al Goalposts at low T]{Aluminum goalpost nano-mechanical devices at low temperatures
}

\author[1]{\fnm{Baptiste} \sur{Alperin}}
\author[1]{\fnm{Ilya} \sur{Golokolenov}}
\author[1]{\fnm{Gwénaëlle} \sur{Julié}}
\author[1]{\fnm{Bruno} \sur{Fernandez}}
\author[1]{\fnm{Andrew} \sur{Fefferman}}
\author*[1]{\fnm{Eddy} \sur{Collin}}\email{eddy.collin@neel.cnrs.fr}

\affil*[1]{\orgdiv{Univ. Grenoble Alpes}, \orgname{Institut N\'eel}, \orgaddress{\street{25 rue des Martyrs}, \city{Grenoble}, \postcode{38042}, \country{France}}}

\abstract{Mechanical objects have been widely used at low temperatures for decades, for various applications; from quantum fluids sensing with vibrating wires or tuning forks, to torsional oscillators for the study of mechanical properties of glasses, and finally micro and nano-mechanical objects with the advent of clean room technologies. These small structures opened up new possibilities to experimentalists, thanks to their small size.
We report on the characterization 
of purely metallic goalpost nano-mechanical structures, which are employed today for both quantum fluids studies (especially quantum turbulence in $^4$He, $^3$He) and intrinsic friction studies (Two-Level-Systems unraveling).
Extending existing literature, 
we demonstrate the analytic modeling of the resonances, in good agreement with numerical simulations, for both first {\it and second} mechanical modes. Especially, the impact {\it of the curvature} of the whole structure (and therefore, in-built surface stress) is analyzed, together with nonlinear properties. We demonstrate that these are of geometrical origin, and {\it device-dependent}. Motion and forces are expressed in meters and Newtons experienced at the level of the goalpost's paddle, for {\it  any} magnitude or curvature, which is of particular importance for quantum fluids and solids studies.
}


\maketitle
\newpage
\section{Introduction}
\label{intro}

Nanomechanical (NEMS) probes enabled within recent years to probe quantum fluids down to very small scales, addressing fundamental questions of condensed matter physics. For instance, nanomechanical strings and goalpost structures have been employed for sensing elementary excitations in superfluid $^4 \mathrm{He}$ \cite{vik4He,Kampi4He}, and recently a suspended string demonstrated interaction with single quantized vortices \cite{vikVortex}.


As well, these probes can also serve the study of (quantum) solids, namely by addressing the mechanical properties of their constitutive elements.
NEMS structures, realised in the clean room, can rather easily be made of (mono-crystalline) silicon, amorphous silicon nitride (SiN), and/or (poly-crystalline) sputtered metals.
In the case of amorphous materials, as studied with paddle torsion oscillators \cite{andrew}, the aim is to understand the nature of the microscopic elements that enable to dissipate mechanical energy: these are believed to be tunneling Two-Level-Systems (TLSs) \cite{philips,anderson}, which in most cases are not identified.

There is also an intriguing mystery: at very low temperatures, poly-crystalline or even mono-crystalline materials have a tendency to display mechanical properties of glasses \cite{kleiman1987two}. The logical conclusion is that all materials contain a certain amount of defects that behave as TLSs, and the question is can we prove it, and ultimately locate them and identify them.
 NEMS devices at low temperatures offer then the unique capability, in principle, to probe only a small group of tunneling TLSs, typically from 10 to 100.

Various nano-mechanical devices have been developed over the years \cite{Bowen, Painter, Kippenberg}. A recent achievement, which is prolongating earlier works on micro-mechanical (MEMS) goalposts \cite{JLTPcollin}, or bilayer NEMS devices \cite{kunal},   
is the purely metallic goalpost geometry of Refs. \cite{Kampi4He,kampidimensional}. This device is made of a single material, here aluminum, which can be either superconducting or normal depending on magnetic field and temperature. It also gives access to large motion, as compared to a doubly clamped geometry. Finally, it also enables to keep the mechanical resonance frequencies rather low (sub-MHz or MHz), which is a requirement for instance in the study of superfluid $^3$He where one would like to keep frequencies much smaller than the superfluid gap, typically 40$~$MHz.

Here we report on the fabrication, modeling and cryogenic characterization of aluminum goalpost structures of thickness about 60$~$nm (denoted "thin") and 150$~$nm ("thick").
Measurements are performed by means of the magnetomotive technique \cite{roukes}. The first and second modes of the structures are presented, with data compared to both 
numerical simulations and analytic expressions \cite{JLTPcollin,theoryED}. The dynamics of the first mode is analyzed taking into account the curvature of the structures, which comes originally from surface-stress in-built during fabrication.
This enables us to properly express the force and the motion experienced by the goalpost. We demonstrate that non-linearities are of the {\it Duffing} type \cite{EddyDuffing}, with no nonlinear damping, and of geometrical origin.  

\section{Results}
\label{results}

\subsection{Design and fabrication}

\begin{figure}[H]
\centering
\includegraphics[width=\textwidth]{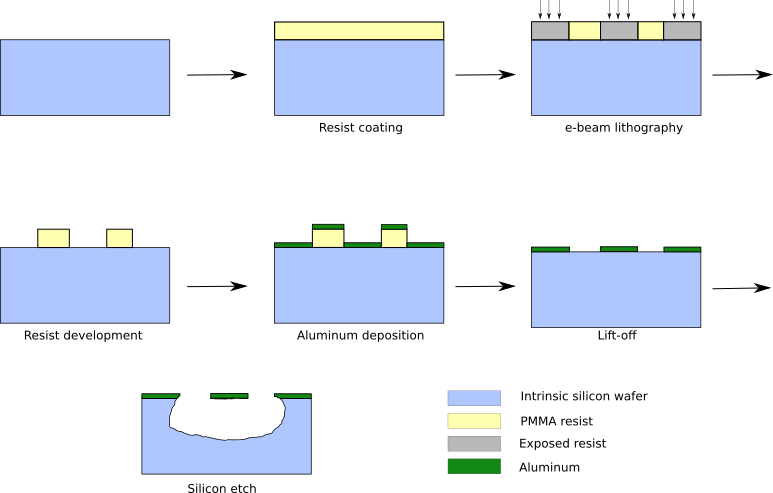}
\caption{Schematics of the fabrication process.
}
\label{fig1}
\end{figure}

Each step of the fabrication process is illustrated in Fig. \ref{fig1}.
The chip used is a 1$~\text{cm}^2$ chip of intrinsic silicon, cut from a larger wafer, protected by a resist layer made of PMMA $4~\%$. After the cutting, we start the process by cleaning the chip from the PMMA using acetone and an ultrasonic bath for 5 minutes. The residues left by the acetone are cleaned by Isopropyl alcohol (IPA), and also an ultrasonic bath. The chip is then rinsed with a RBS solution and then with abundant deionized water before being dried with nitrogen.  \\

Before the 
e-beam lithography, we spin-coated $270~$nm of PMMA $4~\%$ at a speed of $4000~$rpm with an acceleration of $4000~$rpm/s for $60~$s followed by 5 minutes of baking at $180~$°C. 
The contact pads were drawn with a current of 30$~$nA, and dose 700$~\mu$C/cm$^2$ while the resonators were done with a current of 1$~$nA and dose 1000$~\mu$C/cm$^2$.
We made the development using an MIBK:IPA (1:3) solution for $35~$s and then we put the chip in IPA for 1 minute to stop the development of the resist. We cleaned the chip using the same process as before but without the ultrasonic baths that could break the goalpost patterns. \\ 

The chip is cleaned using $\text{O}_2$ plasma for 5 seconds to remove any residual resist. We continue by depositing $60~$nm of aluminum on the sample using e-beam evaporation at a pressure of $5\times 10^{-7}$ Torr. The deposition rate is $0.1~$nm/s. The metal outside of the pattern (deposited on the resist) is ripped off using a lift-off process where the chip is kept in a bath of NMP at $80~$°C for several hours. We typically squeeze the NMP solution on the chip with a pipette directly on the chip. We follow by a classic solvant cleaning with acetone and IPA. 
In another batch, $150~$nm were deposited instead of 60$~$nm. \\

To etch the silicon under the layer of deposited aluminum, we use a dry etch with $\text{XeF}_2$ vapor. The chemical reaction is as follows: 
\begin{equation*}
2\text{XeF}_2(g)+\text{Si}(s)\rightarrow 2\text{Xe}(g)+\text{SiF}_4(g). 
\end{equation*}
The etching occurs as the silicon surface reacts with $\text{XeF}_2$ and forms a gaseous molecule of $\text{SiF}_4$. This non-plasma etching shows an interesting etch selectivity of silicon against other materials such as photoresist, metals or even semiconductor compounds. This allows us to have an isotropic silicon etch to release the suspended structure that we intend to measure. 

\begin{figure}[H]
\centering
\includegraphics[trim={0cm 10cm 4cm 7cm}, clip , width=0.7\textwidth]{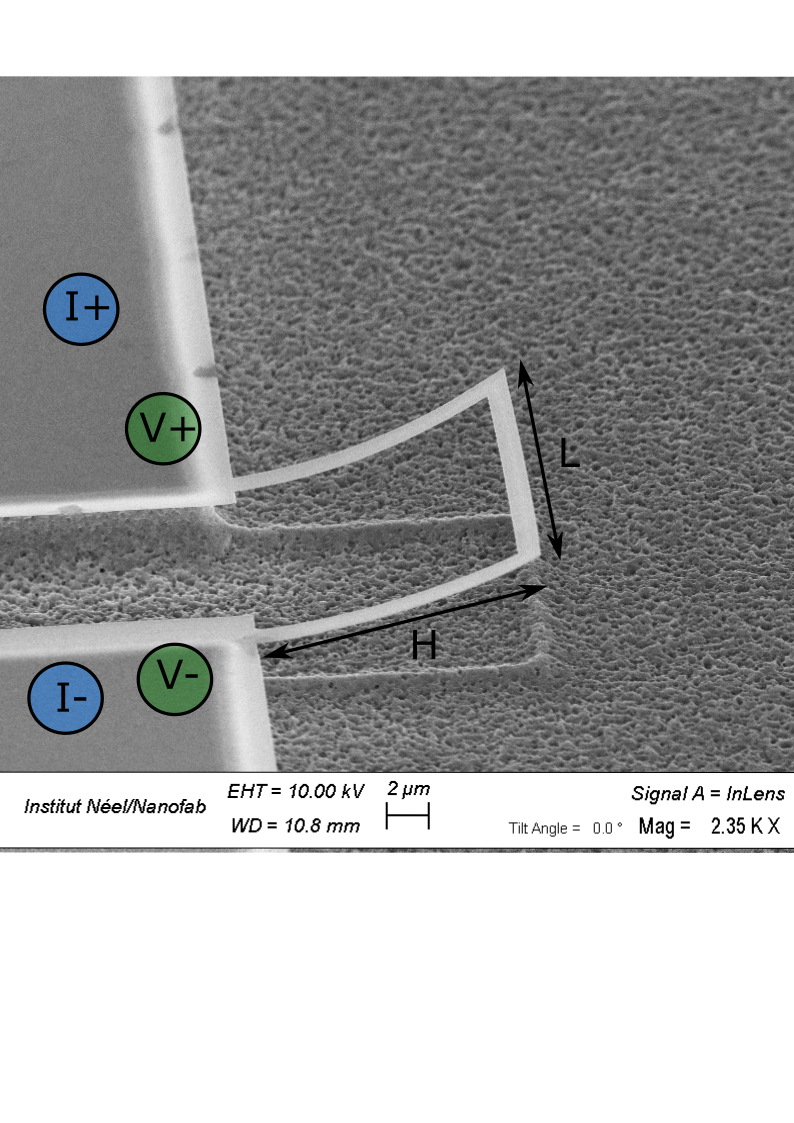}
\caption{Scanning Electron Microscope (SEM) picture of a suspended aluminum resonator with dimensions $H$ and $L$ represented along with the points $I,V$ where the micro-bonding should be made for the 4-point measurement. Note the upwards curvature (see text).
}
\label{fig2}
\end{figure}

Too much etching could cause the structure to collapse. Since the silicon under the clamps is also etched, our goalposts are no longer attached mechanically to anything and thus might touch the substrate beneath. 
If this happens, the structure is stuck to the background and unusable.
\\
The etching process is thus critical for our experiment, and we faced a trade-off problem with the release of our goalposts. 
%
With an RIE etch of the chip by a mixture of $\text{SF}_6$ at $45~$sccm and $\text{O}_2$ at $10~$sccm for 30 minutes at $20~$W, we could etch the silicon enough to suspend the resonators after the $\text{XeF}_2$ etch, as seen in Fig. \ref{fig2}. 
But a by-product of the fabrication is an in-built surface stress that bends the structure upwards. 
As well, due to the difference in thermal contraction between silicon and aluminum, additional thermal stresses will appear as we cool down the devices to Kelvin temperatures. 
These effects can be more or less pronounced, but should definitely be taken into account for the analysis.
The geometrical parameters of our devices are summarized in Tab. \ref{Tablecomsolmgeom}. 
We shall call "feet" the lower parts that link the goalpost to the substrate (of length $H$), and "paddle" the end element (length $L$). The width of all beams is $w$, while the thickness is $e$. Two types of devices have been utilized: so-called "thick" devices (150$~$nm), which are similar to the ones of Ref. \cite{kampidimensional}, and "thin" ones (60$~$nm) which are half the thickness. Presumably, it should be possible to fabricate goalposts reliably down to 30$~$nm, but this has not been tried in the present work and remains to be tested. 

\begin{table}[h!]
\centering
\begin{tabular}{ l ||  c }
  parameter  & value \\
    \hline	
  $ e$ (thickness, target) & 60/150 nm \\
   \hline 
   $w$ (width, measured) &  1.1 µm  \\
   \hline	
   $H$ (length foot, measured) &   13 µm \\
   \hline
   $L$ (length paddle, measured) & 15 µm\\
\end{tabular}
\caption{Geometrical parameters (as measured with the SEM prior to the release, or target value).} 
\label{Tablecomsolmgeom}
\end{table}

\subsection{Numerics and analytics}

We start by presenting COMSOL Multiphysics$^\text{\textregistered}\,$ simulations of our thinnest structures. 
COMSOL is a finite element method simulation tool, which focuses on solving physical problems (based on differential equations, here strain/stress relations) numerically by discretizing the system into small elements \cite{comsol}.
The physical parameters we chose for aluminum are listed in Tab. \ref{Tablecomsolmat}. 
The Young's modulus is fixed to match at best our measurements, within 10$~\%$ of the expected bulk value. We believe that the discrepancy comes from the actual nature of the nanocrystalline thin film.
The modeling considers a single layer of homogeneous, isotropic material, with a clamped boundary condition at the feet's end. 
In these simulations, we did not include any stress, neither from the bulk nor from the surface.
 The simulated structure is thus perfectly flat.
The absence of bulk stress is natural from the free boundary condition at the end of the structure. However, the large curvatures we observe (see e.g. Fig. \ref{fig2}) are necessarily caused by surfacic stresses, which can be regrouped in a mean $\sigma_s^T$ and differential stress $\Delta \sigma_s$ as defined by Sader et al. \cite{Sader2007}.
These terms also are neglected in the simulation, and the relevance of this assumption is discussed further in Section \ref{magnet} below.

\begin{table}[h!]
\centering
\begin{tabular}{ l ||  c }
    & Comsol parameters \\
    \hline	
   Young modulus $E_y$ & 65 GPa \\
   \hline 
   Density $\rho$  &  $2.7\times10^3$ kg/$\text{m}^3$  \\
   \hline	
   Poisson ratio $\nu$  &   0.3 \\
\end{tabular}
\caption{Value of the material constants chosen for the COMSOL simulation.}
\label{Tablecomsolmat}
\end{table}

\begin{figure}[H]
    \centering
    \includegraphics[width=0.52\textwidth]{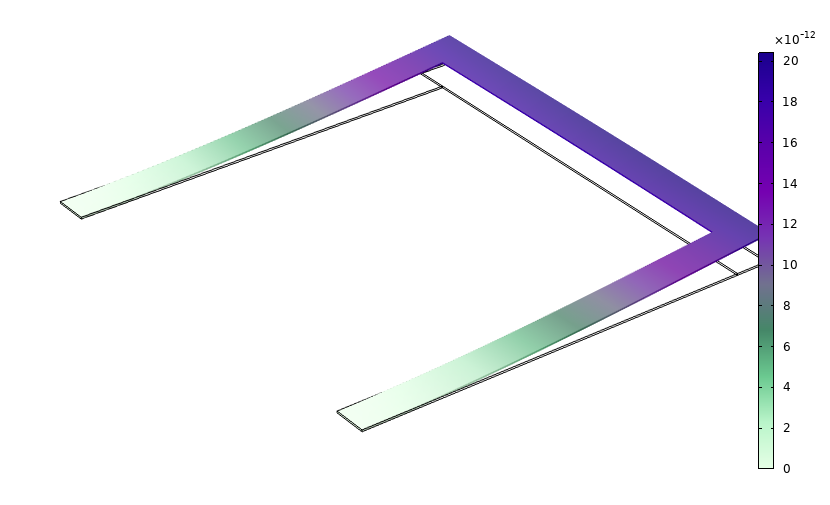} 
    \includegraphics[width=0.47\textwidth]{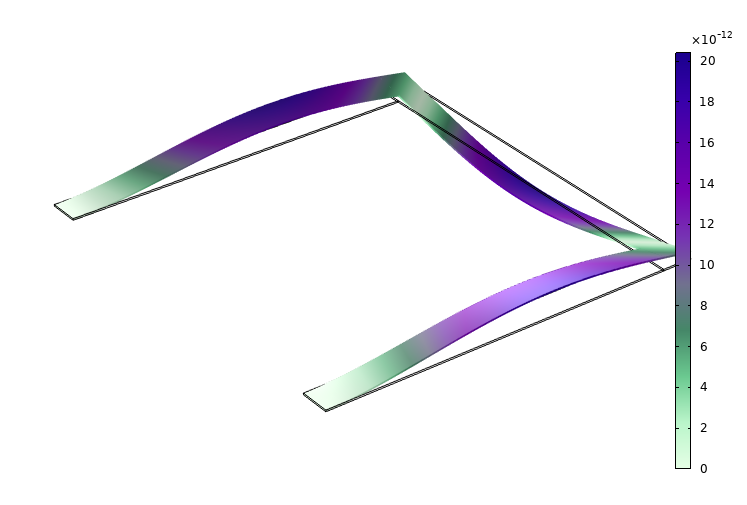} 
    \caption{COMSOL simulation of the first mechanical mode (left), and of the second observable mode of our goalposts (right). 
    The colors show the displacement from the feet plane (arbitrary units). 
    For details, see text.}
\label{fig3}
\end{figure}

One of the COMSOL simulation advantages is to predict the modes (specifically their shapes), even for very complex mechanical structures. 
Examples for the two lowest detectable modes of our thin goalposts are shown in Fig. \ref{fig3}.
We obtain for these frequencies $f_1=153~$kHz and  $f_2=1 095~$kHz, very similar to the analytic description and actual data presented below.
%
However, one of the disadvantages of the COMSOL simulation is that it does not give access directly to the mode parameters (the mode effective mass $m$, and spring constant $k$): in order to compute them, we will have to introduce the analytical description of these resonators.
Furthermore, when one changes a parameter (like e.g. the thickness $e$), the whole simulation has to be restarted all over again.
In this respect, the analytic description is not simply more practical to use, but it also brings additional information: one immediately obtains {\it the dependencies of our parameters on geometrical and physical constants} (for instance, how the frequency $f_1$ depends on the thickness $e$). 
Such a piece of information would require multiple numerical simulations with varying $e$, in order to reconstruct the law $f_1(e)$.

The analytic theoretical description of the first mode of a goalpost resonator can be found in Ref. \cite{JLTPcollin}; the description of higher modes is presented in Ref. \cite{theoryED}.
In the latter reference, the {\it twist} of the beams constituting the structure is taken into account. This makes the modeling substantially more complex, and we shall neglect this aspect here in order to keep our discussion at the most practical level.
We will here briefly remind the essential results required for the modeling, which is based on the Euler-Bernoulli theory \cite{clelandbook}.
We define $I_y=\frac{1}{12}we^3$ the  second moment of area of one foot. 

The shape of the foot under motion is described by a function $f_\text{foot}(x,t) = A(t) \, \Psi_{\text{foot}}(x)$, with $A(t)$ a harmonic motion amplitude parameter (oscillating at frequency $\omega$, 
characterizing the amplitude of motion at the extremity),
 and $\Psi_{\text{foot}}(x)$ the so-called foot mode shape. The boundary conditions are ideal clamp on one end (connection to the substrate), and mass load for the free end (no bending moment, and force load equal to the inertia of the paddle). This leads to:
\begin{equation}
\label{finalpsifoot}
   \!\!  \Psi_{\text{foot}}(x)=A_m \left( [\cos(\lambda x) - \cosh(\lambda x)] - \frac{\cos(\lambda)+\cosh(\lambda)}{\sin(\lambda)+\sinh(\lambda)}[\sin(\lambda x)-\sinh(\lambda x)] \right) , \nonumber
\end{equation}
with $A_m$ an amplitude normalization constant, and $x$ the normalized coordinate along the foot (running from $0$ at clamp to $1$ at paddle).
We chose here $A_m$ such that $\Psi_{\text{foot}}=1$ at $x=1$.
$\lambda$ is the {\it mode parameter} that defines the mode's resonance frequency:
\begin{equation}
\label{freqsval}
\omega_0 = \lambda^2\sqrt{\left( \frac{E_y I_y}{\rho we \, H^4} \right)} ,
\end{equation}
which is obtained from the implicit equation:
\begin{equation}
\label{eqmodalnumber2}
    \frac{\lambda^3+\cosh(\lambda)\left[ \lambda^3 \cos(\lambda)+M \sin(\lambda)\right] -M \cos(\lambda)\sinh(\lambda)}{\lambda^3}=0,
\end{equation}
where we have introduced the normalized mass-loading parameter $M$:
\begin{equation}
    M=-\alpha \, \lambda' \lambda^3.
\end{equation}
\begin{figure}[H]
\centering
\includegraphics[width=0.75\textwidth]{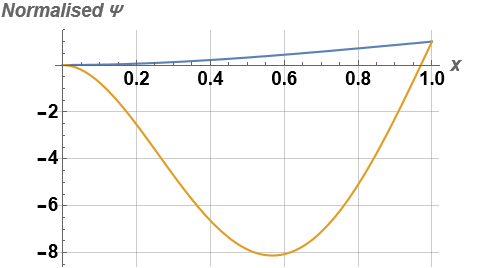}
\caption{Foot mode shape $\Psi_\text{foot}$ for the first mode in blue, and the second one in orange; the normalization is 1 at $x=1$ (see text).
}
\label{fig4}
\end{figure}
The parameter $\alpha=m_{load}/(\rho w eL) $ quantifies how much the paddle actually loads one foot, and $\lambda' = \lambda \, L/H$ will be used below in the calculation of the paddle shape \cite{theoryED}. It is actually this mode-dependent paddle shape that defines self-consistently $\alpha$.
Eq. \ref{eqmodalnumber2} 
is {\it the mode equation} that presents a discrete set of solutions, and which we need to solve in order to find the $\lambda=\lambda_n$ of the $n$-th mode.
We plot in Fig. \ref{fig4} the actual foot shapes of the first two modes of interest $\Psi_\text{foot}(x)$, for the cases $M \approx -2.1$ (first, $\lambda_1$) and $M \approx -610.$ (second, $\lambda_2$). These values of $M$ will be commented below.
Fig. \ref{fig4} can be directly compared to Fig. \ref{fig3}: the visual agreement is reasonably good.

\begin{figure}[H]
    \centering
    \includegraphics[width=0.45\textwidth]{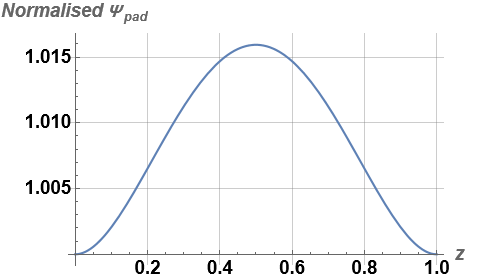} 
    \includegraphics[width=0.45\textwidth]{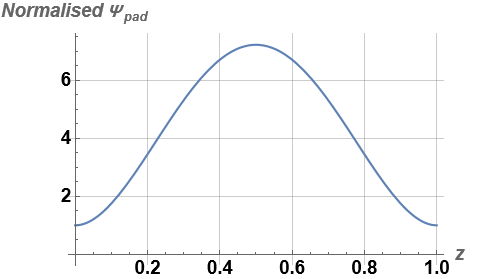} 
    \caption{Paddle mode shapes for the two first modes (see text).}%
\label{fig5}
\end{figure}

As for the foot, we write $f_\text{pad}(z,t) = A(t) \, \Psi_{\text{pad}}(z)$ for the paddle (with $0<z<1$). Solving the Euler-Bernoulli corresponding problem, one obtains \cite{theoryED}:
\begin{equation}
\label{psipad}
    \Psi_{\text{pad}}(z)=\frac{\lambda'\sin(\lambda'/2)\cosh (\lambda'\frac{(1-2z )}{2}) + \lambda'\sinh(\lambda'/2)\cos (\lambda'\frac{(1-2z )}{2})}{\lambda' \sin(\lambda'/2)\cosh(\lambda'/2) + \lambda'\sinh(\lambda'/2)\cos(\lambda'/2)}. \nonumber
\end{equation}
The calculation is represented in Fig. \ref{fig5} for the two first modes,
with $\lambda_1$ obtained solving for a mass loading term $M\approx -2.1$ and $\lambda_2$ for $M\approx -610.$ (as for the feet plot, Fig. \ref{fig4}). 

The analytic modeling is finally completed by defining self-consistently the value of the $\alpha$ parameter.
This is performed by computing the
 bending force exerted by the paddle onto one foot. One obtains \cite{theoryED}:
 \begin{equation}
        \alpha =\frac{2\sin(\lambda'/2)\sinh(\lambda'/2)}{\lambda'[\cosh(\lambda'/2)\sin(\lambda'/2)+\cos(\lambda'/2)\sinh(\lambda'/2)]}.
\end{equation}
For the first mode, we find $\alpha \approx 0.50$ which leads to $M \approx -2.1$. This is a fairly intuitive outcome: half of the paddle mass loads each foot. This results in a net reduction of $\lambda_1$ by a factor about $0.74$. 
For the second mode, we obtain on the contrary $\alpha \approx 2.1$, which leads to $M \approx -611.$ and a reduction factor of about $0.85$.
The mass loading is actually {\it enhanced} by the distortion of the paddle.\\

Let us consider our thin devices (nominal $e=60~$nm). 
Injecting the values we found for $\lambda_1$ and $\lambda_2$ into Eq. \ref{freqsval}, we are then able to find a resonance frequency of $f_1=155.~$kHz for the first mode, and $f_2=1.26~$MHz for the second mode.
These values match well the ones found in the COMSOL simulation (about 15$~\%$ discrepancy at worst here; note that some experimental parameters are not that well known, especially the thickness $e$).
Knowing the functions $\Psi_{\text{foot}}(x)$ and $\Psi_{\text{pad}}(z)$, we can now compute all the required mode parameters. 
The spring constant and mass associated to one foot write:
\begin{eqnarray}
    k_{\text{foot}} & =&  \frac{E_yI_y}{H^3} \int_0^1 \left(\frac{\partial^2 \Psi_{\text{foot}}(x) }{\partial x^2} \right)^2 d x, \\
    m_{\text{foot}} &=& \rho weH \int_0^1 \left(\Psi(x)_{\text{foot}} \right)^2 dx,  \\
    m_{load} & = & \rho weL \, \alpha(\lambda') , \\
     \xi &= & \int_0^1 \Psi_{\text{pad}}(z)dz ,
\end{eqnarray}
with $k=2 k_{foot}$ and $m=2(m_{foot}+m_{load})$ the total spring and mass of the mode verifying $\omega_0=\sqrt{k/m}$. Also, $\xi$ is the {\it mode shape factor} that appears below in the drive/detection scheme.
We obtain for instance $k=5.4 \times 10^{-3}~$N/m and $m=4.3~$pg for the first mode of a "thin" device.

\begin{figure}[H]
    \centering
    \includegraphics [{trim=1cm 4.5cm 1cm 2.2cm},clip, width=0.9\textwidth]{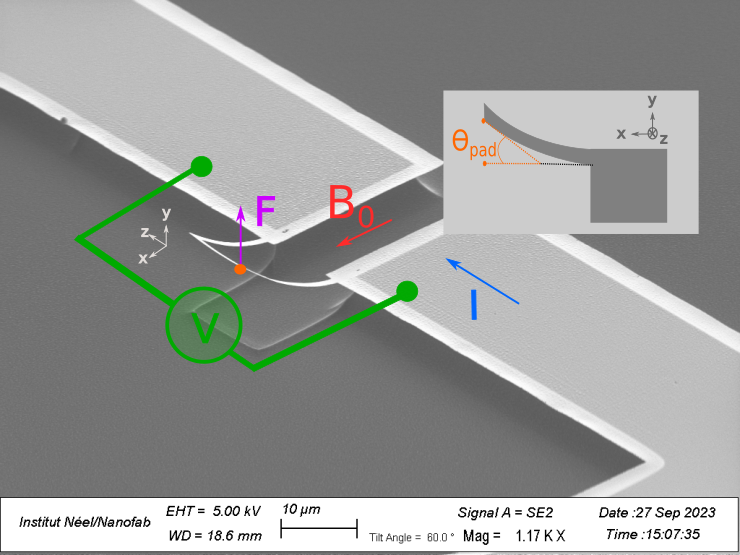}
    \caption{Schematic representation of the magnetomotive actuation and detection principle: $I$ the drive current, $B_0$ the applied field, and $V$ the detected voltage. Note the static curvature of the structure, schematized on the inset with the angle $\Theta_{pad}$ (see text).}
    \label{fig6}
\end{figure}

\subsection{Magnetomotive characterisation}
\label{magnet} 
 
Our measurements are based on the magnetomotive technique \cite{roukes} which we recall rapidly. 
Let us consider first a non-curved structure, that would be contained within a plane parallel to the chip substrate. 
A static magnetic field $B_0$ is applied along the chip, and an oscillatory current $I=I_0 \cos(\omega t)$ with $\omega \approx \omega_0$ is fed in the metallic layer (through a 1$~$k$\Omega$ bias resistance). 
The impedance of the devices is typically about 100$~\Omega$, much smaller than the drive resistor.
A Laplace force is thus exerted onto the paddle:
\begin{equation}
\label{calcforce}
    F_{res}(t) =\xi L I(t) B_0,
\end{equation}
with $\xi$ the shape factor defined above, and $F_0=\xi L I_0 B_0$ the force amplitude. As the structure distorts, the paddle cuts field lines and a voltage appears across the feet (Lenz law):
\begin{equation}
\label{voltsmeters}
V(t) = \xi L B_0 \dot{A}(t).
\end{equation}
One easily verifies the identity: $V(t) I(t) = F(t) \dot{A}(t)$
which states the conservation of power (i.e. energy) for our system. The magnetomotive scheme is depicted in Fig. \ref{fig6} above.
Note that even though the setup is {\it not} 50$~\Omega$ matched, the transmission through the lines is flat up to a few MHz. We therefore do not require a complex calibration procedure as the one presented in Ref. \cite{CollinRSI2012}. \\

But the structures we deal with {\it are usually not} flat.
The distortion seen on the SEM pictures in Figs. \ref{fig2},\ref{fig6} brings in an extra complexity to the problem. The goalpost will indeed oscillate around its rest position, which is making an angle 
$\theta_{pad}$ with the plane of the chip. As such, the actual force $F_{appl}$ acting on the paddle, projected onto this direction, will be obtained through the replacement:
\begin{equation}
    B_0 \rightarrow B_0 \, \cos (\theta_{pad}), \label{angleBfield}
\end{equation}
which has also to be applied to the detected voltage formula. 
On the other hand, the component $\propto \sin (\theta_{pad})$ couples to {\it longitudinal} modes within the feet, which resonate at higher frequencies and are not the subject of our study.
We demonstrate below that from the measurements we can infer the value of the angle $\theta_{pad}$. 
Note that since $\theta_{pad}$ is obtained while the device is cold, it contains both in-built and thermal contraction related contributions.  \\

\begin{table}[h!]
\centering
\begin{tabular}{ l ||  c | c }
    Device and mode & Frequency $f_0$ & Quality factor Q \\
    \hline	
   GP08 ("thin") first & 180$~$kHz & $3\,270.$ \\
   \hline 
    GP08 ("thin") second  &  1.147$~$MHz & $3\,005.$  \\
   \hline  
   GP09 ("thick") first & 388.$~$kHz & $5\,105.$ \\
   \hline 
    GP09 ("thick") second  &  2.75$~$MHz & $5\,500.$  \\
     
\end{tabular}
\caption{Measured parameters$f_0, Q$ at 4$~$K for two typical devices.}
\label{TablesummQ}
\end{table}

\begin{figure}[H]
    \centering
    \includegraphics[width=0.46\textwidth]{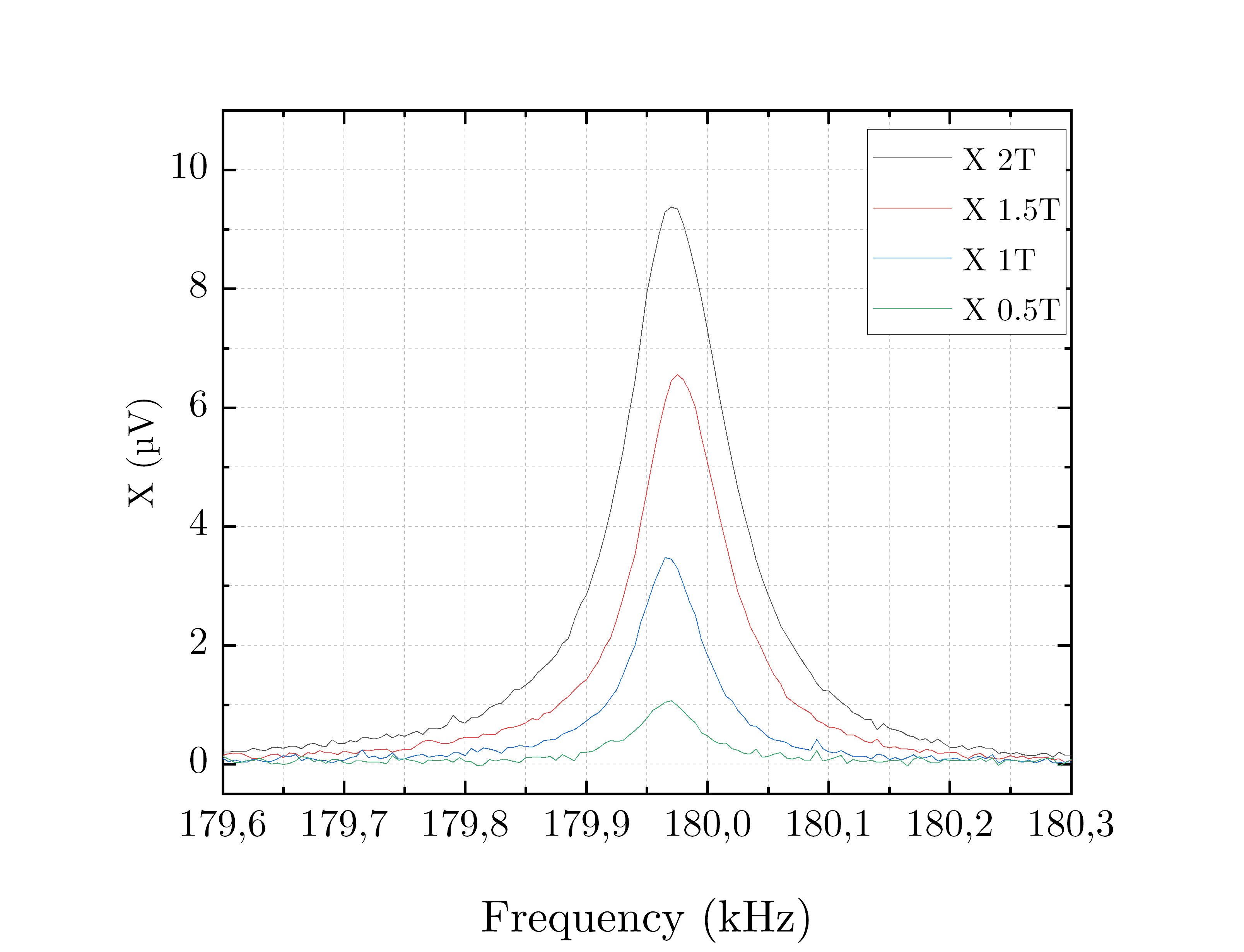} 
    \includegraphics[width=0.46\textwidth]{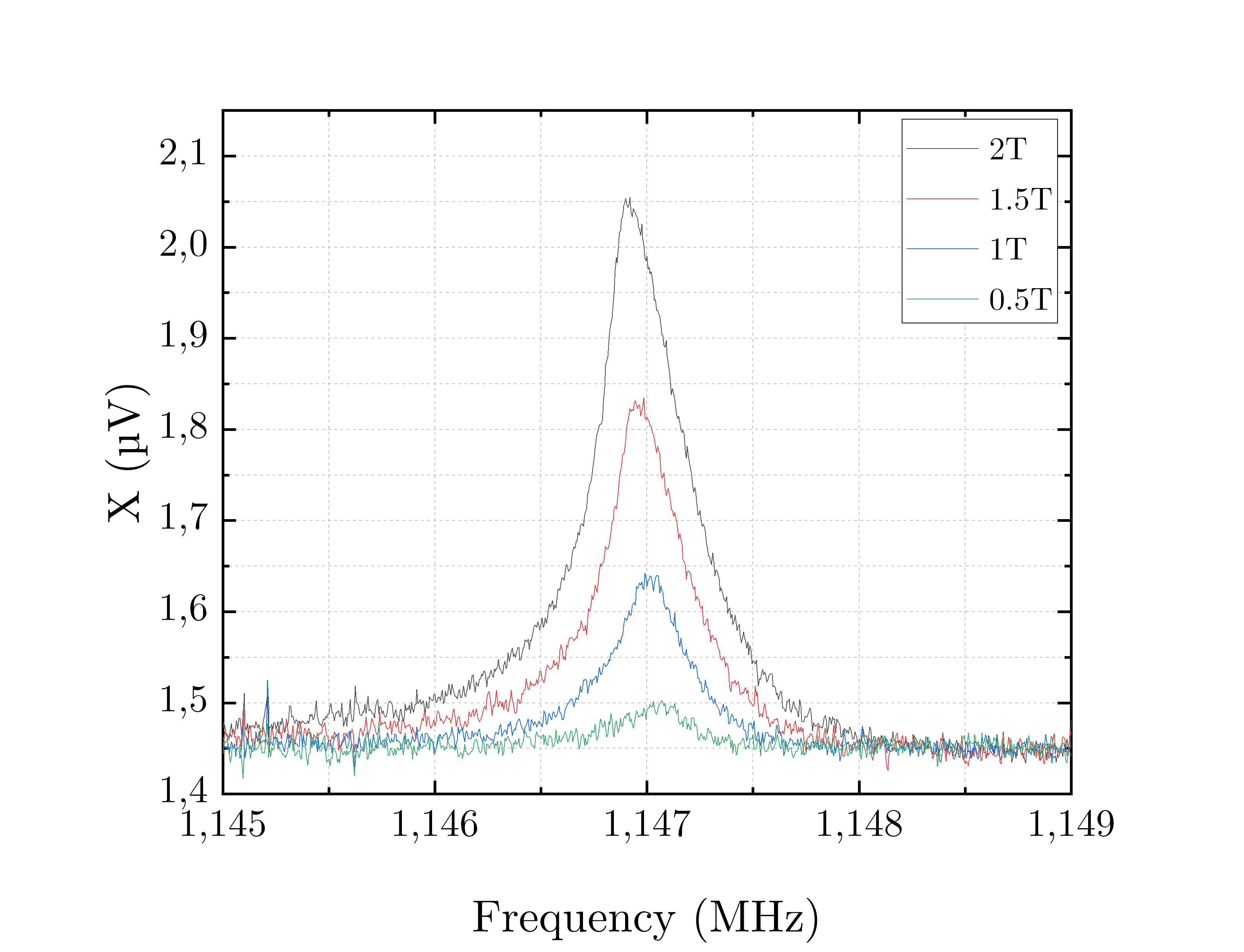} 
    \caption{Measurement of the in-phase $X$ response of the first mode (left) and the second mode (right) of a "thin" device (labeled GP08), while changing the field from 0.5~T to 2~T at 4~K with 70~nA of drive current for the first mode and 200~nA for the second mode (see text).}%
\label{fig7}
\end{figure}

The device is placed in a hermetic cell under vacuum, cooled down to cryogenic temperatures in a 4$~$K cryostat. 
As we sweep the frequency $\omega$ of the drive current, we measure with a lock-in amplifier the resonant response in voltage, $X$ in-phase and $Y$ in-quadrature (with $R=\sqrt{X^2+Y^2}$ the magnitude). In Fig \ref{fig7} we show the resonance peaks obtained in the linear regime, for the two first modes of a "thin" device. The drive force has been kept low enough such that the response is in the linear regime.
The positions of the peaks are in good agreement with the theoretical calculations (about 15$~\%$). As a comparison, our reproducibility in the fabrication is estimated to be about 10$~\%$.
Note also that between cooldowns, the reproducibility of the resonances measured on one device are better than 1$~\%$.
Furthermore, the resonances of one of our "thick" structures are $388.~$kHz and $2.75~$MHz, matching again expectations with similar accuracy (about $15~\%$).
A summary of the results is given in Tab. \ref{TablesummQ}.

\begin{figure}[h]
    \centering
    \includegraphics[width=0.9\textwidth]{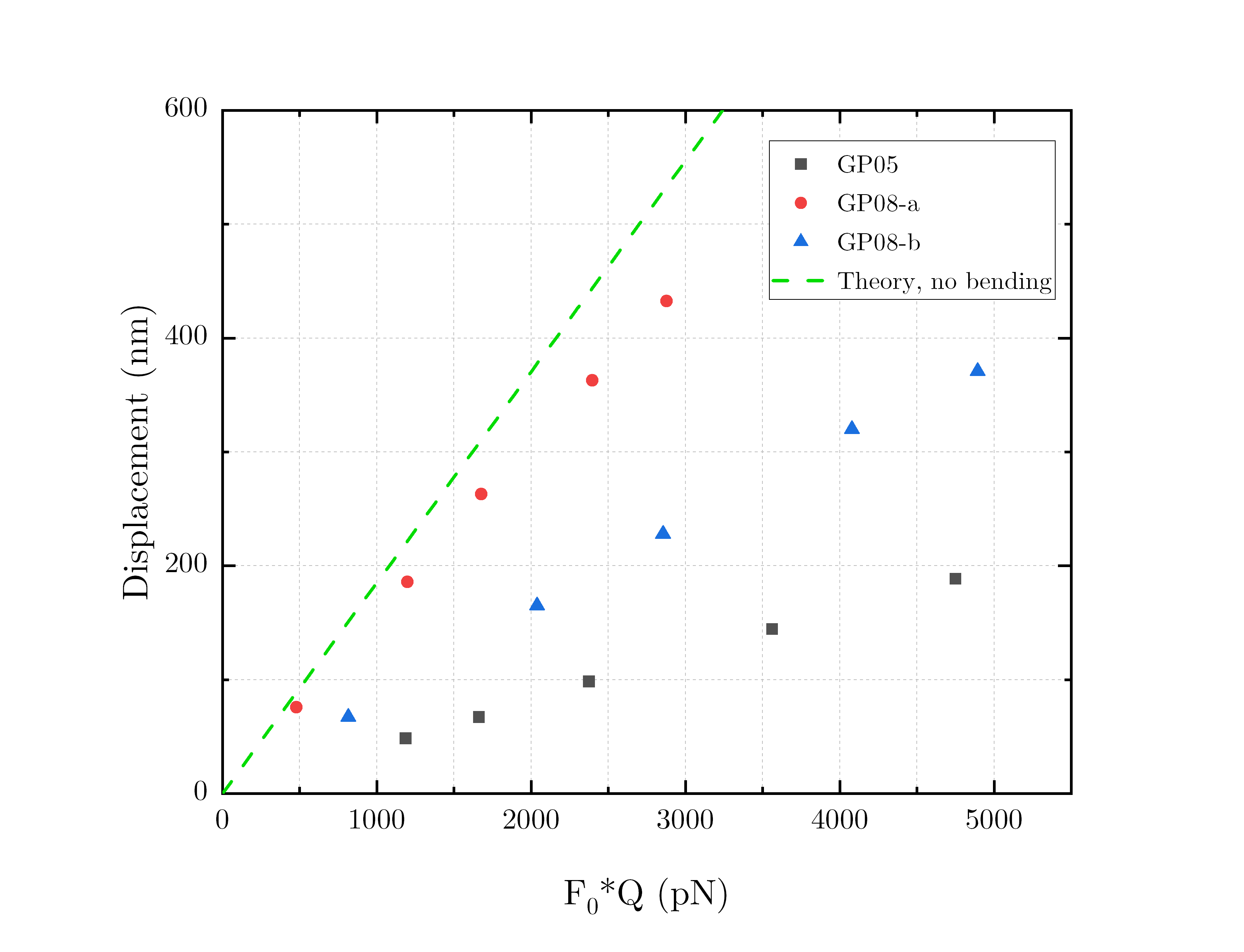}
    \caption{$R_{meas}$ displacement measured on the first mode of two "thin" goalposts, at different drive currents and fields (temperature 4 K), as a function of expected force $F_0$ times $Q$. 
    The dotted line represents the expected $1/k$ slope with no bending. Black dots represent the measurements done on GP05. Blue and red dots represent the measurements of GP08 on two distinct cooldowns (a and b). 
    }
    \label{fig8}
\end{figure}

The linear responses can be fit with Lorentzian expressions, obtaining resonance frequency $f_0$, linewidth $\Delta f$ and maximum height $R_{meas}$, the latter being expressed in meters using Eq. \ref{voltsmeters}. The quality factor is then defined as $Q=f_0/\Delta f$. Knowing the amplitude of the nominal force $F_0$ from Eq. \ref{calcforce}, we can write the ratio:
\begin{equation}
\frac{R_{meas}}{F_0 \, Q} = \frac{R_{max}}{F_{appl}\, Q} \cos (\theta_{pad})^2 = \frac{\cos (\theta_{pad})^2 }{k}, \label{calcAngle}
\end{equation}
which links the detected peak height to $\theta_{pad}$, {\it knowing} the spring constant $k$.
In order to illustrate this, we plot in Fig. \ref{fig8} the measured displacement $R_{meas}$ as a function of the expected force $F_0$ times quality factor $Q$, for the first mode of different "thin" devices (and different runs). The dashed line corresponds to our calculated $1/k$: all data lie {\it below} this line, since $\cos (\theta_{pad})^2 <1$.
 
By fitting the slopes $1/k_{meas}$ in Fig. \ref{fig8}, we can recalculate the bending angle.
This is presented in Tab. \ref{tabTheta}.
The information on the bending angle for each resonator also allows us to recalculate the actual physical displacement $R_{max}$ from the voltage measured. We can thus convert the $X$ and $Y$ responses that we measure in Volts into a proper displacement (in meters).
Note that the angle $\theta_{pad}$ can change substantially from one cooldown to another. 
This clearly demonstrates that the curvature is not only due to the in-built stress created during fabrication: there must also be a contribution from  thermal contraction, which is not always reproducible. \\

\begin{table}[]
\centering
\begin{tabular}{ l || c | c | c }
    
    & $1/k_{meas}$ (nm/nN) & $\theta_{\text{pad}}$ (degrees) & $\text{R}_\text{c}$ ($\mu$m) \\ 
    \hline	
   no bending & 185 & $0^{\circ}$ & $\infty$ \\ 
   GP05 & $40$ & $62^{\circ}$  & 6.9 \\
   GP08-a & $152$ & $25^{\circ}$ & 27.9 \\
   GP08-b & $74$ & $51^{\circ}$ & 10.5 \\
   
\end{tabular}
\caption{Summary of the different angles $\theta_{\text{pad}}$ and the radius of curvature obtained from $ \tan \theta_{pad} \approx \frac{H}{R_c}$, measured on different goalposts with the same thickness. }
\label{tabTheta}
\end{table}

The curvature created in cantilever plates and membranes due to differential surface stress is known as {\it Stoney's problem} \cite{Stoneybending}. 
Following Sader  \cite{Sader2007}, if $R_c$ is the radius of the curvature of the goalpost feet, 
$e$ the thickness of the aluminum film,
$E_y$ its Young's modulus and $\nu$ its Poisson ratio, the Stoney equation reads:
\begin{equation}
\label{stoneyequation}
    \Delta \sigma_s =\frac{E_y \, e^2}{6(1-\nu) \, R_c},
\end{equation}
with $\Delta \sigma_s$ the differential surface stress (in Pa$\times$m). Values of $R_c$ are listed in Tab. \ref{tabTheta}.
 Using Eq. \ref{stoneyequation}, we obtain surface stresses in the range of $3-11$~Pa$\times$m. 
As reminded in Ref. \cite{Sader2007}, differential surface stress $\Delta \sigma_s$ cannot influence the resonance frequencies of our structures.
However, the two surface stresses have no reason to compensate exactly: there certainly is also a small remaining {\it total surface stress} $\sigma_s^T \neq 0$ that might modify the structure's stiffness, and thus modify the resonance frequencies.
In their study, Lachut and Sader \cite{Sader2007} defined for the frequency shifts $\delta \omega$:
\begin{equation}
\label{stressfreq}
    \frac{\delta \omega}{\omega_0}=\phi(\nu) \frac{(1-\nu) \, \sigma_s^T}{E_y \, e}\left(\frac{w}{H} \right) \left(\frac{w}{e}\right)^2.
\end{equation}
\begin{figure}[t]
    \centering
    \includegraphics[width=\textwidth]{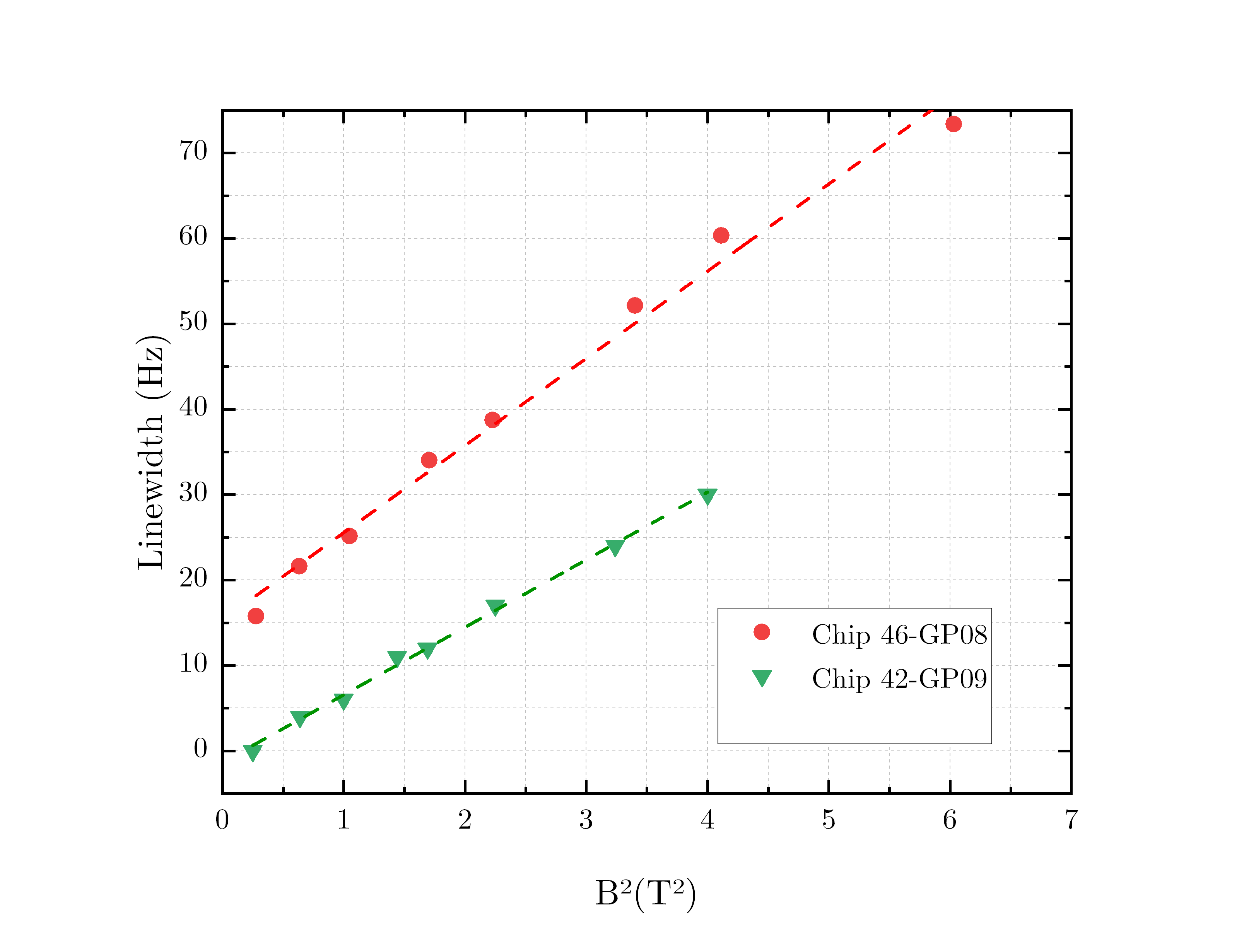}
    \caption{Linewidth of the first mode resonance as a function of the magnetic field squared. Dotted lines represent the linear fits used to extract $R_{ext}$. The data have been shifted for a better visual presentation ( -40$~$Hz for GP08 and -76$~$Hz for GP09). Colors stand for different devices, measured at 4~K with about 80~nA.}
    \label{fig9}
\end{figure}
They found from numerical simulations that for rectangular thin-and-long cantilevers ($H \gg w \gg e$), the function $\phi(\nu) \propto \nu$. For the first mode, they give explicitly $\phi(\nu) \approx - 0.042 \, \nu$.
Making a "worst estimate", we can assume that all the differential stress comes from one side of the structure, leading to $\sigma_s^T \approx \Delta \sigma_s$.
Injecting this estimate of $\Delta \sigma_s$ into Eq. \ref{stressfreq}, we obtain typically $\delta \omega / \omega_0 \approx -3 \times 10^{-4}$, much smaller than the typical reproducibility of our measurements (about 10$~\%$).
We therefore validate the hypothesis of neglecting surface stress in our modeling, in both the numerical simulations and the analytics.

\begin{figure}[t]
    \centering
    \includegraphics[width=\textwidth]{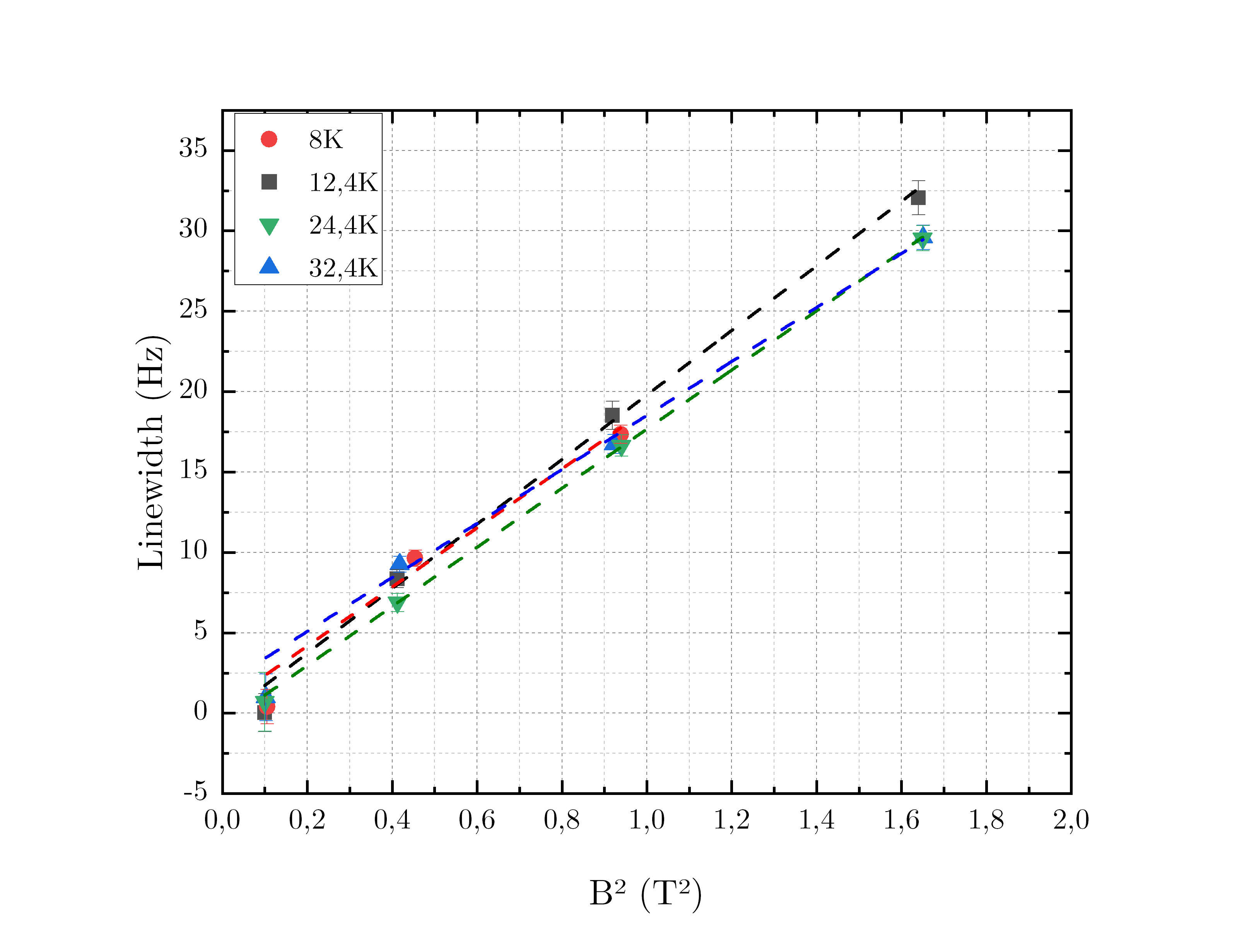}
    \caption{Linewidth of the first mode resonance as a function of the magnetic field squared for different temperatures, measured on goalpost GP08. The data have been shifted for a better visual presentation. Dotted lines represent the linear fits, demonstrating a rather small dispersion (see legend for color code, and text for details).}
    \label{fig10}
\end{figure}

The magnetomotive scheme comes along with an extra damping named magnetomotive loading \cite{roukes}. It is due to the back-action of the external circuit onto the device: the induced voltage being responsible for a drive current generated by the circuit impedance $Z_{ext} \approx R_{ext}$. 
The dependence of the resonance linewidth on the magnetic field is then quadratic:
\begin{equation} 
      \Delta f_{meas} = \Delta f_0 +\frac{ \xi^2 L^2}{ 2\pi \,  m} \frac{ B_0^2 }{  R_{ext}^2} ,  
      \label{eqload}
\end{equation}
with $\Delta f_0$ the intrinsic damping. Note that $B_0$ here should also be renormalized by the bending angle, Eq. \ref{angleBfield}. This one is inferred from the fit $1/k_{meas}$, as reported above. 
We present in Fig. \ref{fig9} the linewidth measured at $4~$K for two different devices (one "thin", and one "thick"). As well, we show in Fig. \ref{fig10} the temperature dependence measured on one of them. 
We confirm that the dependence is indeed $\propto B_0^2$, in all sets of data. As well, we see that the loading is temperature-independent in the range $4 -30~$K (typically within $\pm 10~\%$): this is consistent with a constant bending angle $ \theta_{pad} $, and a constant circuit impedance $R_{ext}$.
From Eq. \ref{eqload}, we recalculate the load impedance; the results are summarized in Tab. \ref{tabexternalimpedance}. 
The obtained values are close to the nominal imposed bias resistance (about a k$\Omega$), regardless of the device (its thickness, or the mode frequency). On the other hand, it is {\it very different} from what is calculated from the eddy current argument proposed in Ref. \cite{kampidimensional}: we would then obtain an effective load impedance of about $1~\Omega$. We therefore conclude that the eddy current model, as proposed by these Authors, does not apply to our devices. \\

\begin{table}[h!]
\centering
\begin{tabular}{ l || c | c || c |c || c }
    
    & GP09 (150~nm) & Angle & GP08 (60~nm) & Angle & Average $R_{ext} $ \\
    \hline	
   4~K cryostat & 1.2~k$\Omega$ & $47^{\circ}$ & 1.9~k$\Omega$ & $51^{\circ}$ & 1.5~k$\Omega$\\
\end{tabular}
\caption{Summary of the calculated external impedance, using  two different goalposts. See text for discussion. }
\label{tabexternalimpedance}
\end{table}

\begin{figure}[H]
    \centering
    \includegraphics[width=0.49\textwidth]{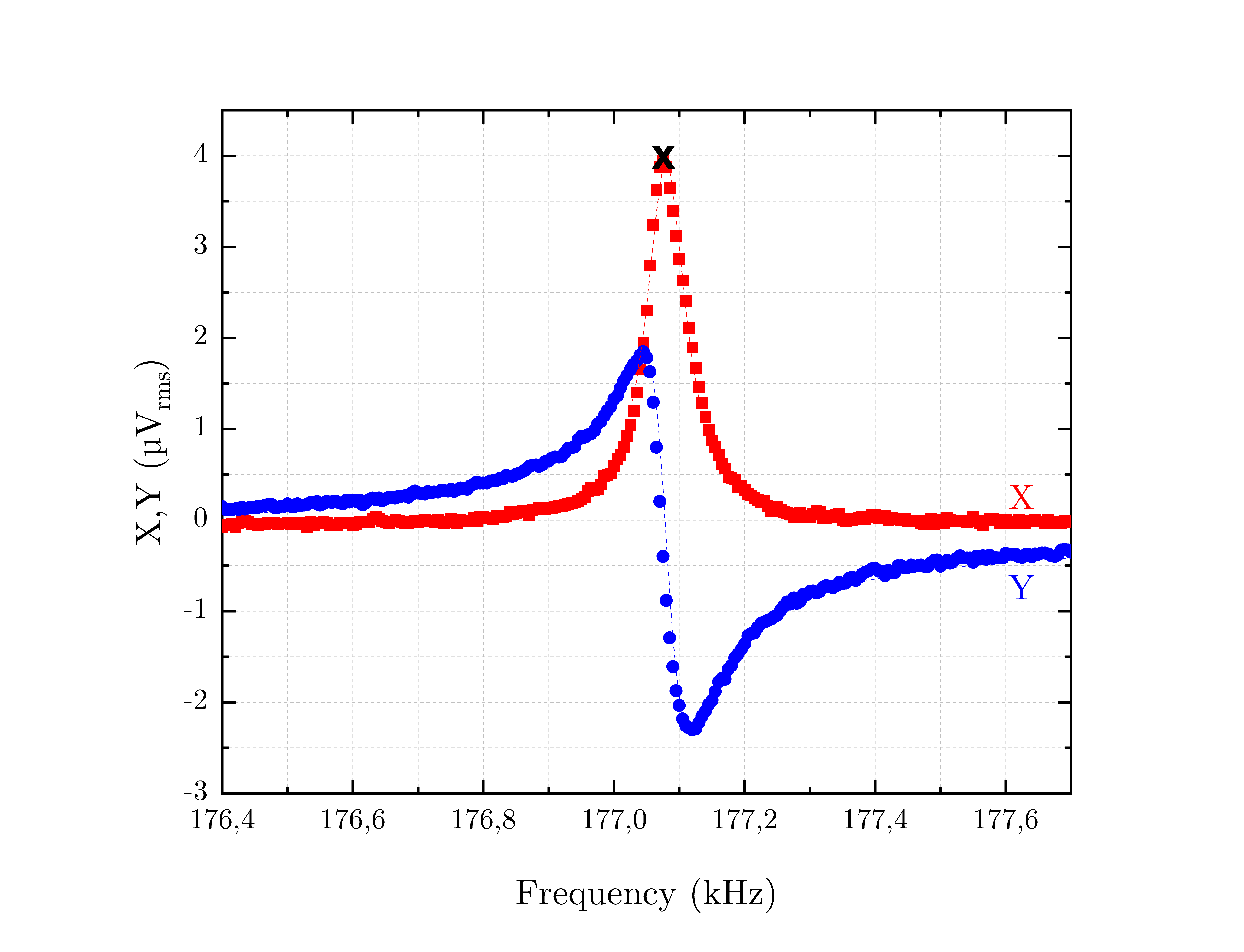} 
    \includegraphics[width=0.49\textwidth]{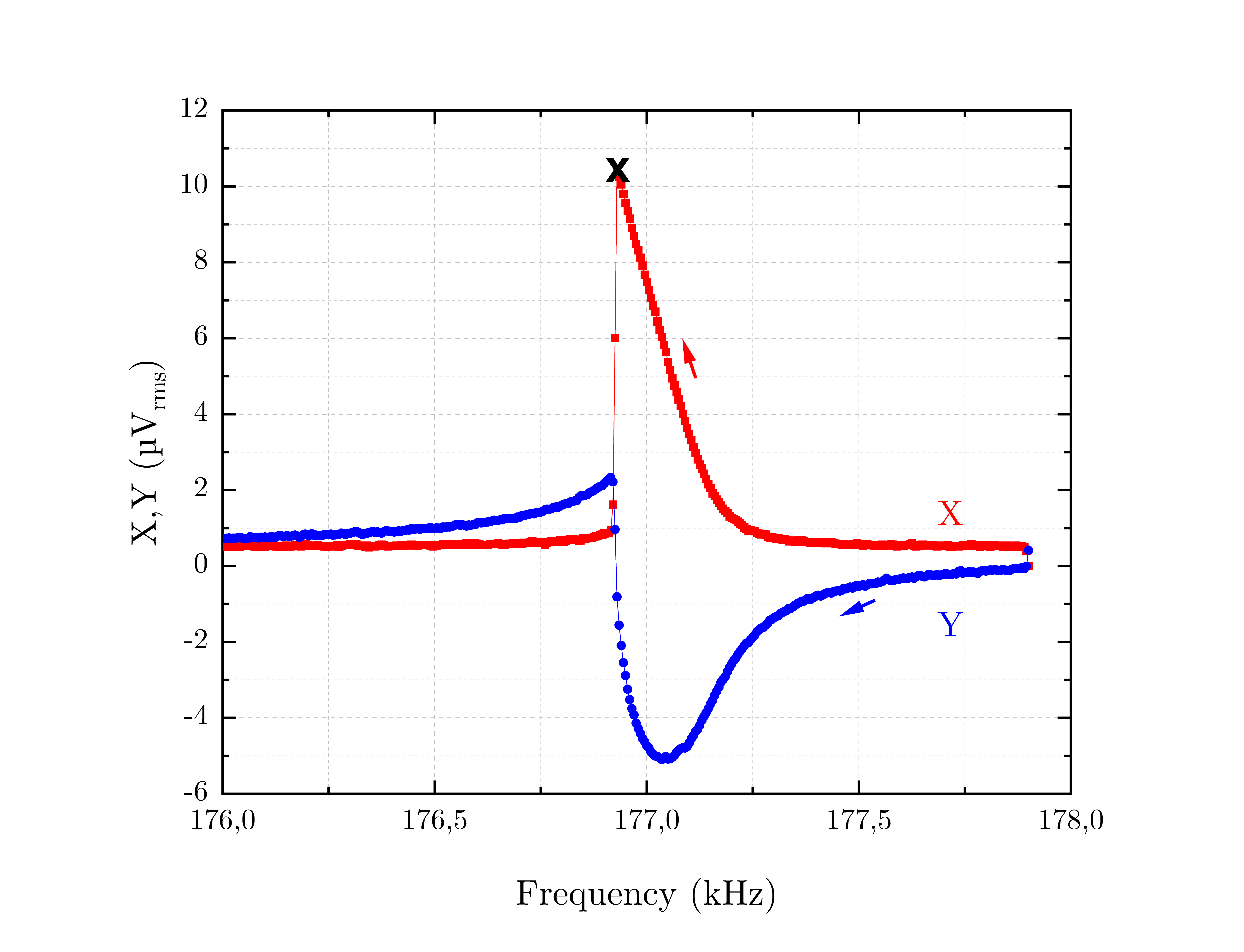} 
    \caption{A "Thin" device resonance measured at 4~K (in vacuum). Left, linear (Lorentzian) regime: Measurement done with 50~nA of drive current, with a magnetic field of 1.2~T.  Right, nonlinear (Duffing) regime, 
   same device with drive current 200~nA and field 1~T. 
   The black cross defines position (in Hz) and maximum amplitude (here in Volts measured prior to corrections, see text).}%
\label{fig11}
\end{figure}

As we increase the driving force, the mechanical resonance will eventually display nonlinear features. Strikingly, it tilts and displays a characteristic Duffing-like peak \cite{EddyDuffing}, as illustrated in Fig. \ref{fig11}.
It turns out that if there is no other source of nonlinearities such as {\it nonlinear damping}, the maximum height still verifies the law $R_{max}=F_{appl} Q/k$ and the frequency shifts quadratically with amplitude $\delta f = \beta R_{max}^2$ (downwards for a negative Duffing coefficient $\beta$).

\begin{figure}[h]
    \centering
    \includegraphics[width=\textwidth]{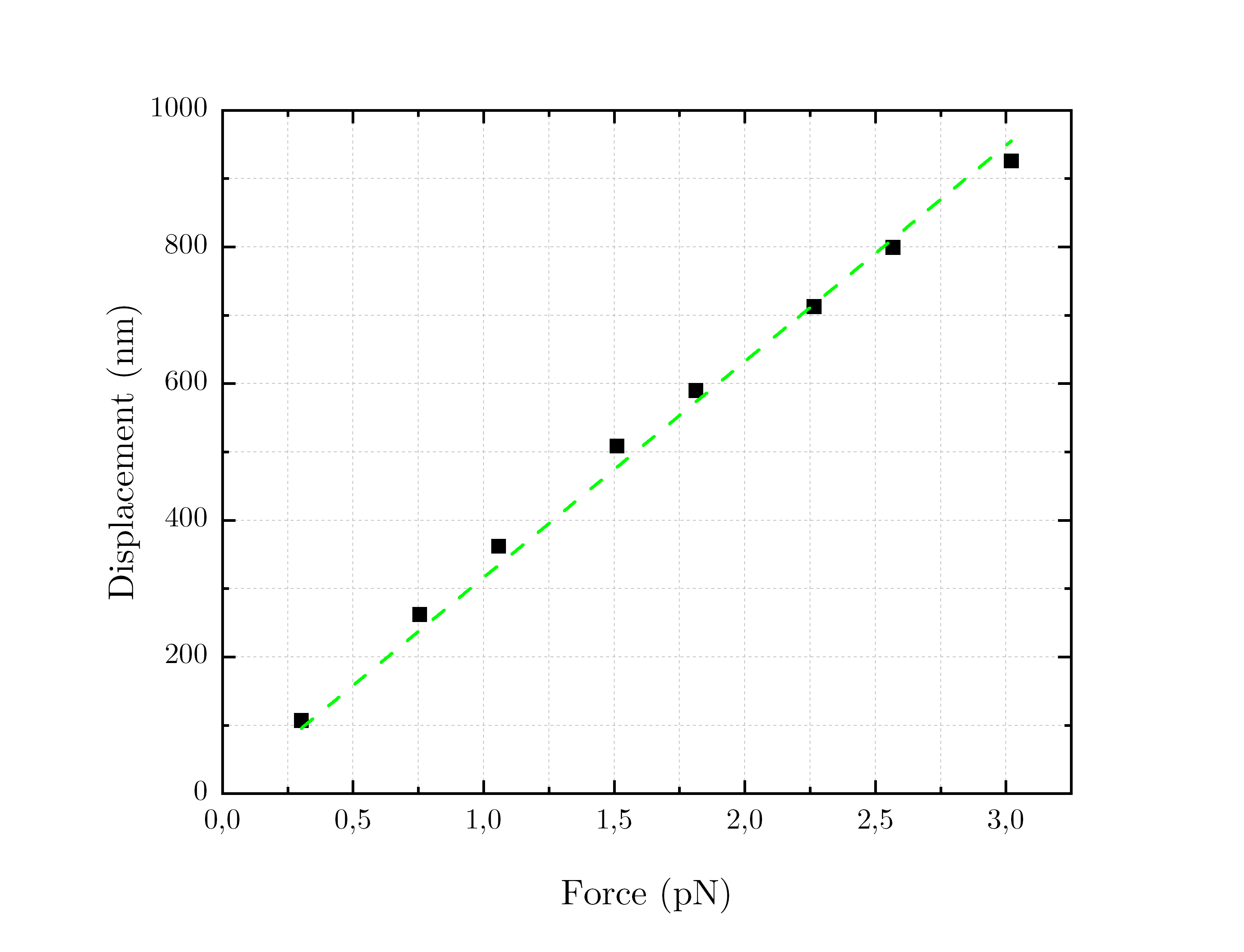}
       \caption{Displacement vs force (device GP08) measured at $4~$K in $1~$T at different currents. The proper height of motion is obtained through Eq. \ref{calcAngle} with the calculated bending angle $\theta =51^{\circ}$. In green is the linear fit $\propto Q/k$ (see text).
    }
    \label{fig12}
\end{figure}

\begin{figure}[h]
    \centering
    \includegraphics[width=\textwidth]{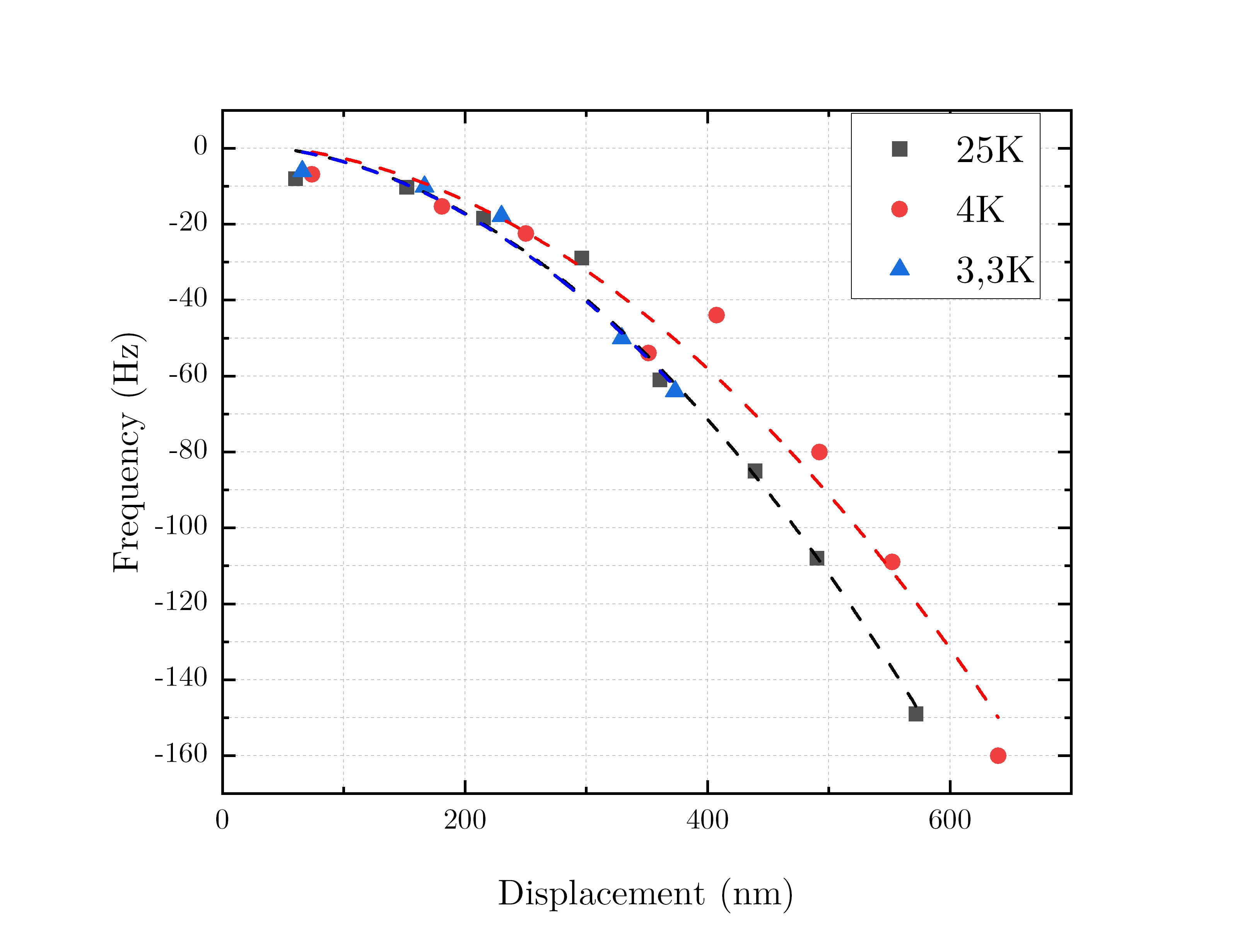}
    \caption{Duffing frequency shift as a function of displacement for the "thin" device GP08, measured at different temperatures.  
    The displacement has been renormalized with a bending angle $\theta=25^{\circ}$. 
    The dotted lines represent the fits used for the extraction of $\beta$ (see text for details).}
    \label{fig13}  
\end{figure}

This can be analyzed with the position of the black cross in Fig. \ref{fig11} in both amplitude and frequency. We first plot in Fig. \ref{fig12} the height versus force. 
The expected linear fit indeed holds up to very large motion amplitudes (almost 1$~\mu$m), demonstrating the absence of nonlinear damping. 
Note that this is not always the case, and seems to depend mostly on the materials used \cite{CollinJAP2010}.
We can thus plot in Fig. \ref{fig13} the frequency shifts, measured at different temperatures. The data at 3.3$~$K has been obtained by pumping on the liquid $^4$He bath of the cryostat. The different quadratic fits illustrate the dispersion in our data, about $\pm 15~\%$. As for the loading effect, the measurements are consistent with a bending angle $\theta_{\text{pad}}$ independent of temperature, and also a {\it constant Duffing parameter} $\beta$.  This is consistent with a nonlinearity of geometrical origin \cite{EddyDuffing}.

\begin{figure}[h]
    \centering
    \includegraphics[width=\textwidth]{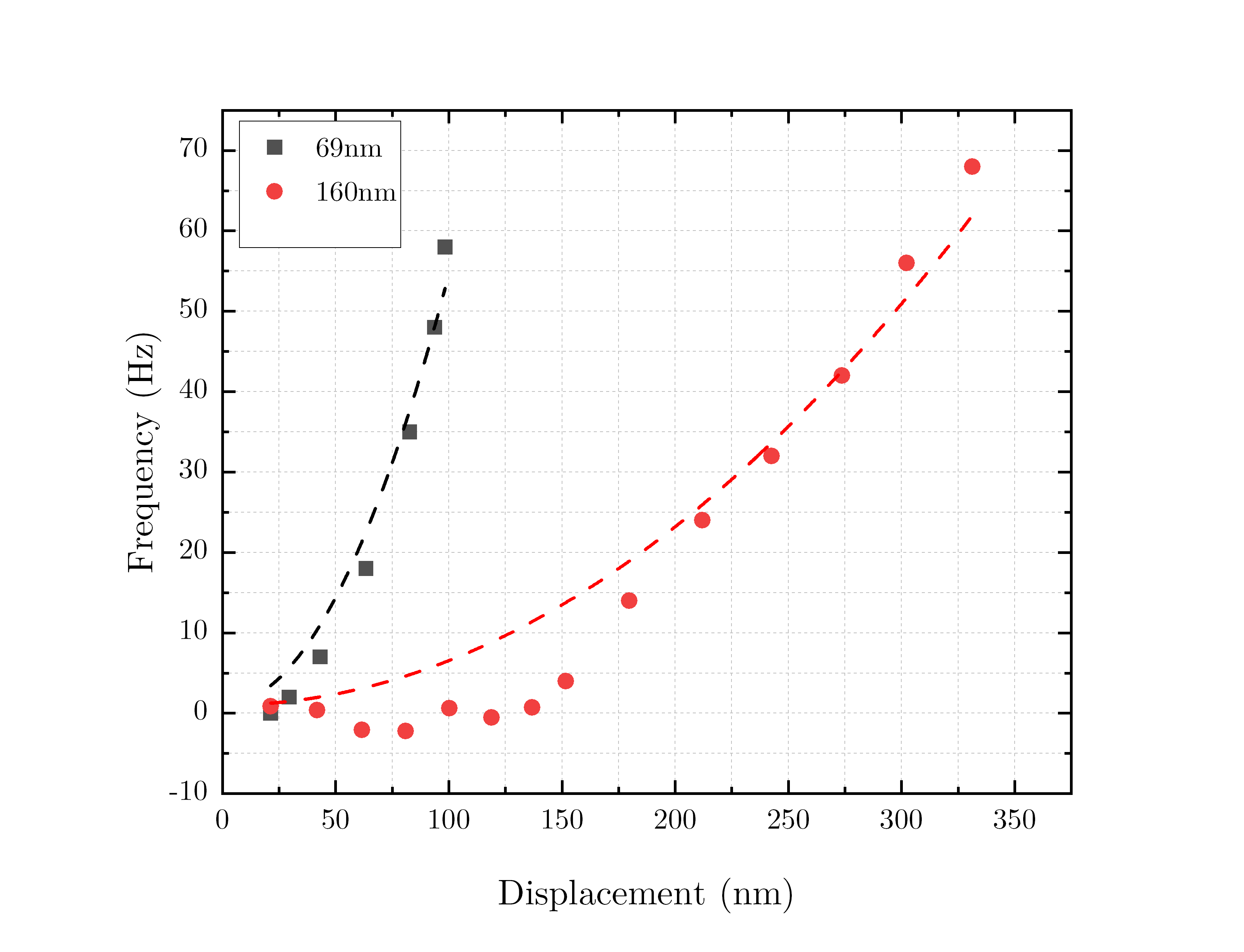}
    \caption{Frequency shift as a function of the displacement for GP05 (a "thin" device, 60~nm) and GP09 ("thick", 150~nm) at 4~K. 
    The displacement has been renormalized with the calculated bending angle for each structure ($64^{\circ}$ and $47^{\circ}$ respectively). The dotted lines represent the fits used for the extraction of $\beta$, see text.}
    \label{fig14}
\end{figure}

However, what actually defines this Duffing coefficient remains mysterious. 
The GP08 Duffing coefficient was measured in two cooldowns with different bendings (about $25^{\circ}$ and $45^{\circ}$), and  the inferred values agree within $\pm 30\%$, which is about the best we could expect for such measurements in terms of reproducibility, taking into account all possible sources of errors. 
Therefore, the Duffing nonlinearity does not seem to depend on bending, or surface stress.
Furhtermore, in Fig.  \ref{fig14}, we present similar data measured for another "thin" device, and a "thick" one. 
Surprisingly, while for GP08 we had $\beta <0$ we measured here two {\it quite different positive coefficients}.
Understanding why GP08 and GP05 have Duffing coefficients of opposite signs, while being so similar, would require further studies. Neither the bending, nor the thickness $e$ seem to be alone responsible for the effect. 
But this   
 could be of great interest for engineering, since 
 it means that we could then {\it tune the resonator} to be very highly linear with $\beta \approx 0$.
 Even at extreme drives, the resonance would then remain Lorentizan, an extremely interesting feature when one aims at measuring very precisely resonance frequency and/or $Q$ factor.

\section{Conclusion}
\label{concl}

In the present article we have reported the design, modeling and characterization at cryogenic temperatures of goalpost NEMS devices made of aluminum. 
Numerics and analytic expressions have been presented, reproducing well the measured results. The second mode resonance frequencies have been reported, demonstrating good linear properties and quality factor (the $Q$ is typically of the order of a few thousands at 4$~$K). 
Measurements have been performed using the magnetomotive technique.
We took specifically care of the curvature of the structures in our analysis, demonstrating that one can infer the bending angle of the goalposts, and thus recalculate motion and forces in proper meters and Newtons.
This analysis which has been performed here with the first mode, should apply equally well to any other one, especially the second mode of flexure we reported.
The magnetomotive loading has been carefully analyzed, ruling out the eddy current argument proposed in the literature for these structures. 
Finally, we report on the nonlinear properties of the first mode. 
The resonances all look like reasonable Duffing peaks, with no other nonlinearities (especially, no nonlinear damping).
It turns out that while it seems to be of geometrical origin (temperature-independent), the parameters that define the Duffing coefficient are not understood. 
It is nonetheless rather interesting to pursue investigations since we demonstrate both positive and negative Duffing parameters, for rather similar structures.
This essentially means that, if one would master the underlying reasons for this, it could then be possible to tune the nonlinearity to almost zero, guaranteeing linearity up to extremely large amplitudes. This would certainly be of great utility, for instance in the sensing of quantum fluids and of mechanical materials.

\bmhead{Acknowledgments}
We acknowledge the use of the N\'eel Nanofab facility, and we thank very much the  N\'eel Cryogenics team for their help.
The authors acknowledge the support from ERC StG grant UNIGLASS No. 714692.
The research leading to these results has received funding from the European Union's Horizon 2020 Research and Innovation Programme, under grant agreement No. 824109, the European Microkelvin Platform (EMP).


\begin{thebibliography}{9}

\bibitem{vik4He} Probing superfluid $^{4}\mathrm{He}$ with high-frequency nanomechanical resonators down to millikelvin temperatures, A. M.  Gu\'enault, A. Guthrie, R. P. Haley, S. Kafanov, Yu. A. Pashkin, G. R. Pickett, M. Poole, R. Schanen, V. Tsepelin, D. E. Zmeev, E. Collin, O. Maillet, R. Gazizulin, Phys. Rev. B Vol. 100, 020506 (2019).

\bibitem{Kampi4He} Superfluid $^4$He as a rigorous test bench for different damping models in nanoelectromechanical resonators, T. Kamppinen, J. T. Mäkinen, V. B. Eltsov, Physical Review B Vol.  107, 014502  (2023).

\bibitem{vikVortex} Nanoscale real-time detection of quantum vortices at millikelvin temperatures, Guthrie, A. and Kafanov, S. and Noble, M.T. and  Pashkin, Yu. A. and Pickett, G. R. and Tsepelin, V. and Dorofeev, A. A. and Krupenin, V. A. and Presnov, D. E., Nature Comm. Vol. 12, 2645 (2021).

\bibitem{andrew} Elastic measurements of amorphous silicon films at mK temperatures, A. Fefferman, A. Maldonado, E. Collin, X. Liu, T. Metcalf and G. Jernigan, J. of Low Temp. Phys. Vol. 187, pp. 654 (2017).

\bibitem{philips} Tunneling states in amorphous solids, W. A. Phillips, J. of Low Temp. Phys. Vol. 7, pp. 1573-7357 (1972).

\bibitem{anderson} Anomalous low-temperature thermal properties of glasses and spin glasses, P. W.  Anderson, B. I.   Halperin,  C. M.   Varma, The Philosophical Magazine: A Journal of Theoretical Experimental and Applied Physics Vol. 25, pp. 1-9 (1972). 

\bibitem{kleiman1987two} Two-level systems observed in the mechanical properties of single-crystal silicon at low temperatures, R.N. Kleiman, G. Agnolet, D.J. Bishop, Phys. Rev. Lett. Vol. 59, pp. 2079 (1987).



\bibitem{Bowen} Laser cooling and control of excitations in superfluid helium, G.I. Harris,  D. L. McAuslan, E. Sheridan, Y. Sachkou, C. Baker, W. P. Bowen, Nat. Phys. 12, pages 788–793 (2016)
\bibitem{Painter} Nano-acoustic resonator with ultralong phonon lifetime, Gregory S. MacCabe, Hengjiang Ren, Jie Luo, Justin D. Cohen, Hengyun Zhou, Alp Sipahigil, Mohammad Mirhosseini, Oskar Painter, Science Vol. 370, Issue 6518, pp. 840-843 (2020).
\bibitem{Kippenberg} A squeezed mechanical oscillator with millisecond quantum decoherence, Amir Youssefi, Shingo Kono, Mahdi Chegnizadeh, Tobias J. Kippenberg, Nat. Phys. 19, 1697–1702 (2023).



\bibitem{JLTPcollin} Silicon vibrating wires at low temperatures, Eddy Collin, Laure Filleau, Thierry Fournier,  Yuriy Bunkov, Henri Godfrin, J. of Low Temp. Phys. Vol. 150, 739-790 (2008).

\bibitem{kunal} Evidence for the role of normal-state electrons in nanoelectromechanical damping mechanisms at very low temperatures, K.J. Lulla, M. Defoort, C. Blanc, O.  Bourgeois, E. Collin, Phys. Rev. Lett. Vol. 110, 177206 (2013).

\bibitem{kampidimensional} Dimensional control of tunneling two-level systems in nanoelectromechanical resonators, T. Kamppinen, J.T. M{\"a}kinen, V.B. Eltsov, Phys. Rev. B Vol. 105, 035409 (2022). 

\bibitem{roukes} External control of dissipation in a nanometer-scale radiofrequency mechanical resonator, A.N. Cleland, M.L. Roukes, Sensors and Actuators A: Physical Vol. 72, 256-261 (1999).

\bibitem{CollinRSI2012} In-situ comprehensive calibration of a tri-port nano-electro-mechanical device, 
E. Collin, M. Defoort, K. Lulla, T. Moutonet, J.-S. Heron, O. Bourgeois, Yu. M. Bunkov, H. Godfrin, 
Rev. Sci. Instrum. Vol. 83, 045005 (2012).


\bibitem{theoryED} Modal Decomposition in Goalpost Micro/nano Electro-mechanical Devices, E. Collin, M. Defoort, K.J. Lulla, J. Guidi, S. Dufresnes, H. Godfrin, J. of Low Temp. Phys. Vol. 175, 442-448 (2014).

\bibitem{EddyDuffing} Addressing geometric non-linearities with cantilever MEMS: beyond the Duffing model, E. Collin, Yu.M. Bunkov, H. Godfrin, Phys. Rev. B Vol. 82, 235416 (2010).

\bibitem{comsol} COMSOL Multiphysics$^\text{\textregistered}\,$ website: https://www.comsol.com/comsol-multiphysics.

\bibitem{CollinJAP2010} Metallic coatings of microelectromechanical structures at low temperatures: Stress, elasticity, and nonlinear dissipation, 
E. Collin, J. Kofler, S. Lakhloufi, S. Pairis, Yu. M. Bunkov, and H. Godfrin, Journal of Applied Physics, Vol. 107, Issue 11, 114905 (2010).


\bibitem{Sader2007}  Effect of Surface Stress on the Stiffness of Cantilever Plates, Michael J. Lachut, John E.  Sader, Phys. Rev. Lett. Vol. 99, 206102 (2007).

\bibitem{clelandbook} Foundations of Nanomechanics: From Solid-State Theory to Device Applications, A. N. Cleland, Springer Science \& Business Media (2013).

\bibitem{Stoneybending} The tension of metallic films deposited by electrolysis,  George Gerald Stoney, Proceedings of the Royal Society of London. Series A, Containing Papers of a Mathematical and Physical Character Vol. 82, 172-175 (1909).
 
\end{thebibliography}

\bibliographystyle{ieeetran}

\end{document}